\documentclass{elsart}
\usepackage{epsfig}
\usepackage[verbose]{wrapfig}
\graphicspath{{/home/giulia/curioni/NIM_2002/tex/}
       	      {/home/giulia/curioni/NIM_2002/figures/}}
\newcommand{\g}{$\gamma$}

\newcommand{\yt}{$^{88}$Y}

\newcommand{\bc}{\begin{center}}
\newcommand{\ec}{\end{center}}
\newcommand{\be}{\begin{equation}}
\newcommand{\ee}{\end{equation}}
\newcommand{\bfg}{\begin{figure}}
\newcommand{\efg}{\end{figure}}
\newcommand{\bi}{\begin{itemize}}
\newcommand{\ei}{\end{itemize}}
\newcommand{\bt}{\begin{table}}
\newcommand{\enta}{\end{table}}
\newcommand{\keV}{\mbox{ke\hspace{-0.1em}V}}
\newcommand{\MeV}{\mbox{Me\hspace{-0.1em}V}}

\renewcommand{\deg}{\ensuremath{^\circ}}

\newcommand{\version}{June 17, 2003 \enspace (Draft Version 1.1)} 
\begin{document}
\begin{frontmatter}
\title{ Detection of partial polarization in Compton scattered photons }
\date{\version}
\author{A.~Curioni$^a$, E.~Aprile$^a$}
\address{$^a$Columbia Astrophysics Laboratory, Columbia University, New York,
NY, USA}  

\begin{abstract}
   
It has been recently proposed \cite{SEBoggs:nim03} to use polarization of
Compton scattered \g-rays to improve the imaging performance of Compton
telescopes. Building upon that work, we detected the aforementioned
polarization in a sample of 1.836~\MeV\ \g-rays from the LXeGRIT Compton
telescope. Here we present the measurement, together with detector oriented
considerations on the application of the principle to a realistic Compton
telescope.  

\end{abstract}

\end{frontmatter}

\section*{Introduction}

In a recent work \cite{SEBoggs:nim03} it has been shown how polarization of
Compton scattered \g-rays retains information about the source location, and it
has been suggested that it provides a very significant improvement in the
reconstruction (imaging) of a \g-ray source, the most common imaging technique
being based on Maximum Likelihood algorithms (see for example
Ref.~\cite{HdeBoer:92}). \\ 
The basic idea can be outlined as follows: given an unpolarized \g-ray source,
let's assume that two Compton scatters followed by full absorption are
detected. The first scattered photon will be partially polarized in a direction
perpendicular to the scatter plane; the second scattered photon will be
preferentially in a plane perpendicular to the polarization direction, i.e. in
the plane containing  the source, the first interaction site and the second
interaction site. \\ 
It is clear that a fine grained detector is needed in order to be able to detect
each interaction. LXeGRIT \footnote{See Ref.~\cite{EAprile:2002.spie} for a
recent review; a more detailed description is in preparation} is a good example
of such a detector and polarization has been detected in a sample of 3-site
events from 1.836~\MeV\ \g-rays. The measurement is described in detail in
Sec.~\ref{sec:1}. We did not try to assess directly the impact of polarization
in constraining the source position, since it would require a major effort to be
implemented in the current LXeGRIT imaging procedure. We try, anyway, to clarify
what is needed to make this approach effective in Sec.~\ref{sec:2}. \\ 
When coming down to realistic detectors, the field of Compton telescopes is
nowadays rather fluid, without much agreement within the community about the
detector of choice. The sheer lack of such a {\it canon} makes it difficult to
bring even bright ideas from the realm of wishful thinking to practical
realization. To give an example, it was proposed \cite{TJONeil:AIP00} to track
the recoil electron in the first Compton scatter, which is extremely troublesome
since tracking $\sim$\MeV\ electrons requires a low density, low $Z$ active
medium, clearly conflicting with the requirement of stopping power needed for
\g-rays. A realistic design to realize such an idea is not available to date.

\section{\label{sec:1} Measurement}

\begin{figure}[htb]
\centering
\includegraphics[bbllx=55,bblly=505,bburx=550,bbury=730,
        width=\linewidth,clip]{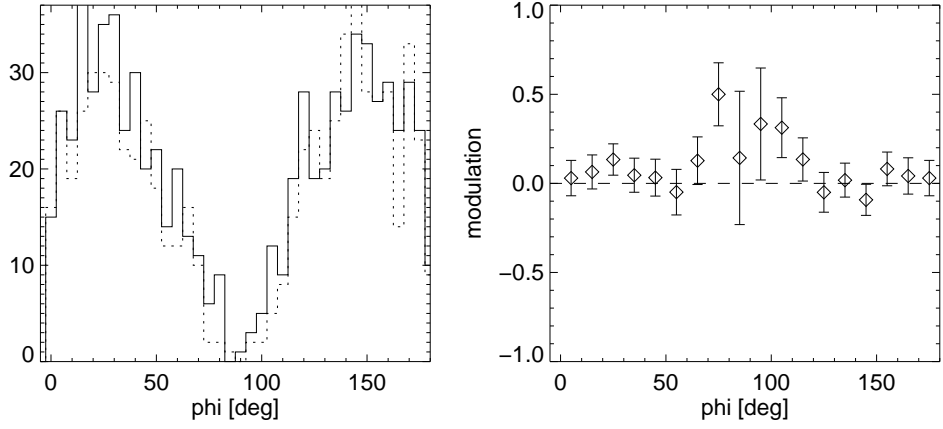}
\caption{ 
{\it Left:} $\phi$ distributions for the two sub-samples $66\deg < \theta _{123}
< 114\deg$ (continuous line) and $\theta _{123} > 114\deg$ or $\theta
_{123} < 66\deg$ (dashed line). {\it Right:} modulation vs. $\phi$.
}
\label{fig:1}
\end{figure}

For a concise and clear exposition of the general problem, we refer to
\cite{SEBoggs:nim03}. \\
LXeGRIT is a Compton telescope which exploits a liquid xenon time projection
chamber (LXe TPC) to detect \MeV\ \g-rays, imaging each interaction in the
fiducial volume with an accuracy better than 1~mm on each coordinate; quite
naturally the TPC provides a built-in Cartesian reference frame. 
Data from an \yt\ \g-source, which emits photons at 0.898 and 1.836~\MeV, have
been used  for this measurement. The source was placed on top of the TPC, at a
distance of 2~m along the $\hat{z}$ axis and centered at $x$=0 and $y$=0 in the
TPC reference frame. The 1.836~\MeV\ line was selected for it provided a better
detection efficiency. 
Thus we experimentally deal with 
\begin{itemize}
\item [1. ]direction of \g-rays coming from the source, which, in the TPC
reference frame is (0,0,1);
\item [2. ] $E_i$, $x_i$, $y_i$, $z_i$ for the $i^{th}$ interaction, $i$=1,2,3;
\item [3. ] $E_1$+$E_2$+$E_3$=1.836~\MeV, because we select events in the full
energy peak. 
\end{itemize}
For each event, the three interactions have been sequenced using Compton
kinematics \cite{UOberlack:2000.spie} and only events which give the correct
source location are then selected; in this way the contamination due to
incorrectly sequenced events is reduced to a truly negligible level. \\
In order to study the polarization of the scattered \g-rays, we change to the
reference frame described in \cite{SEBoggs:nim03}, i.e. the $\hat{z}$ axis in
the direction from the second interaction to the first one and the
$\hat{y}-\hat{z}$ plane defined by the three interaction sites. 
In the new reference frame the vector from the first to the second interaction
is $\vec{v}_{12} \equiv (0,0,-1)$ and the vector from the second to the third
interaction is $\vec{v}_{23} \equiv (0,\sin{\theta _{123}},\cos{\theta_{123}})$;
$\theta _{123}$ is the second scatter angle.
The $\hat{x}$ axis of the new reference frame is now given in the TPC reference
frame as the cross product 
\begin{equation}
\hat{x} = \vec{v}_{12}~' \times \vec{v}_{23}~' 
\label{eq:1}
\end{equation}
properly normalized; primed vectors are in the TPC reference frame. \\
The source position, in the new reference frame, is defined as (Eq.~5 in
\cite{SEBoggs:nim03}) 
\begin{equation}
r_0 = (\sin{\theta _{012}}\cos{\phi},\sin{\theta
_{012}}\sin{\phi},\cos{\theta _{012}})  
\label{eq:2}
\end{equation}
where $\theta _{012}$ is the first scatter angle and $\phi$ is the azimuthal
angle in the $\hat{x} - \hat{y}$ plane; $\phi$ is the sensitive variable in this
measurement.
The $\hat{x}$ component of $r_0$ is found projecting the source direction
(0,0,1) on the $\hat{x}$ axis as defined in Eq.~\ref{eq:1}, and the $\hat{y}$
component then follows.  
Once $r_{0x}$ has been determined, $\cos{\phi}$ is given and we can look for
an excess in the $\phi$ distribution, favoring the plane containing the
source, the first interaction site and the second interaction site
(012-plane). Here we shift $\phi$ of 90\deg, so that the 012-plane
corresponds to $\phi$=90\deg, while in \cite{SEBoggs:nim03} it corresponds to
$\phi$=0\deg,180\deg. 
As shown in \cite{SEBoggs:nim03}, this excess in the $\phi$ distribution
is expected to be small for 1.836~\MeV\ \g-rays; as expected in polarization
phenomena (and confirmed in Fig.~2 in \cite{SEBoggs:nim03}), it is
supposed to be maximum for $\theta _{123} \sim 90\deg$ and negligible for
$\theta _{123} < 30\deg$ and $\theta _{123} > 150\deg$. 
To emphasize the excess in the $\phi$ distribution, we divide our sample in two
sub-samples: one with $66\deg < \theta _{123} < 114\deg$ and the other one with
$\theta _{123} > 114\deg$ or $\theta _{123} < 66\deg$ (the precise choice of
these numbers is not critical). 
Our expectation is now to detect polarization in the first sub-sample while the
level of polarization should be negligible in the second one. The $\phi$
distribution for the two sub-samples is shown in Fig.~\ref{fig:1}-$left$.
We define a $modulation$ as 
\begin{equation}
modulation = \frac{ \bar{\phi _1} - \bar{\phi _2} }{ \bar{\phi _1} + \bar{\phi
_2} }  
\label{eq:3}
\end{equation}
where $\bar{\phi}$ is the $\phi$ distribution binned, in this case, in
intervals of 10\deg; the subscripts indicate the sub-sample. The $modulation$ is
shown in Fig.~\ref{fig:1}-$right$, error bars from statistical errors; a
significant excess is visible in the region $\phi = 90\deg \pm 30\deg$, as
expected.  

\begin{figure}[htb]
\centering
\includegraphics[bbllx=55,bblly=505,bburx=550,bbury=730,
        width=\linewidth,clip]{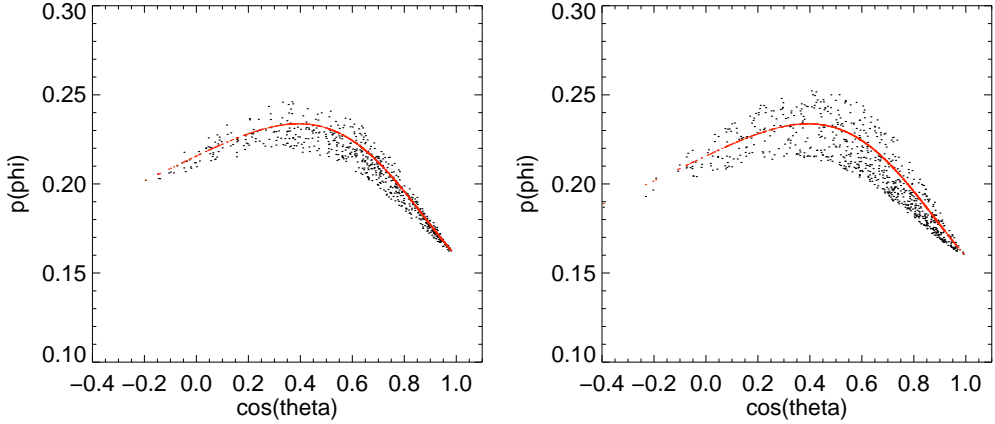}
\caption{ $p(\phi,\theta _{012})$ vs. $\cos{ \theta _{012} }$ compared to
$\tilde{p}(\theta _{012})$ vs. $\cos{ \theta _{012} }$ (continuous line). {\it
Left:} sub-sample 2, {\it right:} sub-sample 1.
}
\label{fig:2}
\end{figure}

\begin{figure}[htb]
\centering
\includegraphics[bbllx=55,bblly=505,bburx=350,bbury=730,
        width=0.65\linewidth,clip]{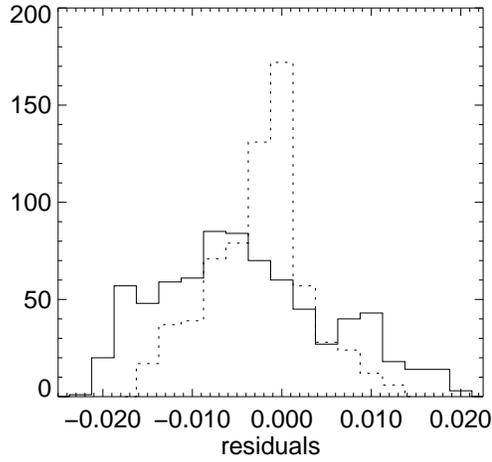}
\caption{ $p(\phi,\theta _{012})~-~\tilde{p}(\theta _{012})$, i.e. residuals
from Fig.~\ref{fig:2}; the {\it dashed line} is for sub-sample 2, the {\it
continuous line} for sub-sample 1.
}
\label{fig:3}
\end{figure}

We can look at the problem in a slightly different way, taking Eq.~10 in
\cite{SEBoggs:nim03} as our starting point; it can be rewritten as 
\begin{equation}
p(\phi,\theta _{012}) = \frac{ \beta + \beta^{-1} - 2 \sin^2 \theta _{012} (I'_1
- (I'_1 - I'_2)\cos^2 \phi )}{2\pi( \beta + \beta^{-1} - 2 \sin^2 \theta_{012}
)}   
\label{eq:4}
\end{equation}
where $\beta$ is the ratio of the photon energy after the scatter to its
energy before scattering, i.e. 
$$
\beta = \frac{0.511}{ 0.511 + E_{\gamma} ( 1 - cos{\theta _{012}} )}
$$
$E_{\gamma}$ = 1.836~\MeV\ in the present case;
$$
I'_1 = \frac{ \beta' + \beta'^{-1} }{2 ( \beta' + \beta'^{-1} - \sin^2
\theta_{123} )}
$$
$$
I'_2 = \frac{ \beta' + \beta'^{-1} - 2 \sin^2 \theta_{123} }{2 ( \beta' +
\beta'^{-1} - \sin^2 \theta_{123} )}
$$
where primed quantities refer to the scattered \g-ray and $E_{\gamma}$=$E_1 +
E_2$ in $\beta'$; here $\phi$ is defined as in \cite{SEBoggs:nim03}. \\  
For unpolarized photons $I'_1 = I'_2 = 0.5$ and Eq.~\ref{eq:4} is reduced to 
\begin{equation}
\tilde{p}(\theta _{012}) = \frac{ \beta + \beta^{-1} - \sin^2 \theta _{012}
}{2\pi ( \beta + \beta^{-1} - 2 \sin^2 \theta _{012} )}  
\label{eq:5}
\end{equation}
which does not depend on $\phi$, as expected. We can therefore compare
$p(\phi,\theta _{012})$ to $\tilde{p}(\theta _{012})$ and look for some
discrepancy. Again, it is advisable to split the data in two sub-samples and
pick up the differences between them. The results for the aforementioned
sub-samples are shown in Figs.~\ref{fig:2},~\ref{fig:3}. Fig.~\ref{fig:2}-$left$
shows $p(\phi,\theta _{012})$ vs. $\cos{ \theta _{012} }$ for the supposedly
unpolarized sub-sample (2), Fig.~\ref{fig:2}-$right$ for the supposedly
polarized sub-sample (1); $\tilde{p}(\theta _{012})$, obviously the same for the
two different samples, has been superimposed (continuous line). In an attempt to
quantify how much discrepancy from the unpolarized case is in the two
sub-samples, the histograms of the residuals are shown in Fig.~\ref{fig:3}; the
{\it dashed line} is for sub-sample 2, the {\it continuous line} for
sub-sample 1. 
The supposedly polarized sub-sample clearly shows a more
pronounced discrepancy from the unpolarized case and its residual distribution
is shifted toward negative values, as expected.


\section{\label{sec:2} Discussion}

We have presented the detection of polarization in a sample of Compton scattered
\g-rays, following the suggestion in \cite{SEBoggs:nim03}. It was done for a
relatively high energy (1.836~\MeV), where the effect is expected to be small,
because it offered the larger sample. \\
To quantify the {\it modulation} as defined in \cite{SEBoggs:nim03} is well
beyond the scope of this work, as well as to quantify its impact on source
imaging. A rough but solid estimate can be given answering the following
question: which fraction of {\it imaging events} \footnote{For an imaging event,
full containment and at least two interactions are required.} ends up in the
{\it polarized sub-sample}, as defined earlier? \\
In LXeGRIT, a large fraction of imaging events is detected with two interactions
only, and for this 2-site events no polarization of the scattered \g-ray is
detectable, as obvious; at 1.836~\MeV, 2-site events make up $\sim$80\% of the
imaging events. 2-site events include a large fraction of the events with a
large ($\sim$90\deg or larger) scatter angle, which are most favorable to detect
polarization. Of the remaining 3-site events, about 50\% are in the {\it
polarized sub-sample}, i.e. $\sim$10\% of the total. This figure gives a
pre-factor that should be included when quoting the modulation along the event
circle; the ratio 3-site events to imaging events is further reduced at lower
energies.  

It is clear at this point that polarization of Compton scattered \g-rays is
a small effect for a detector like LXeGRIT, and this conclusion can be easily
extrapolated to other detectors using dense, high $Z$ active media (e.g. Ge) and
a mm position accuracy. On the other side, polarization has been detected
using a detector by no means optimized to do so, possibly indicating that to use
it to constrain the event circle in Compton imaging is within reach of a
realistic Compton telescope. \\
The main limitation in LXeGRIT does not come neither from the relatively
high energy threshold (150~\keV) or from its mm spatial accuracy (very hard to
improve in any realistic, sizeable \g-ray detector), but from LXe itself, which
has a large cross-section for photoelectric absorption due to $Z_{Xe} = 54$,
exceeding the Compton cross-section below 300~\keV. LXe has a density of
$\sim$3~g/cc and an attenuation length well below 1~cm for photon energies of
200~\keV\ or less, so that many interactions are too close to be spatially
resolved. These properties, which allow to build an extremely compact and
efficient Compton telescope out of LXe, pose a significant limit on the event
multiplicity in LXe, when a multiplicity~$\geq$~3 is required for detecting
polarization of Compton scattered \g-rays. \\
Without abandoning the terrain of noble liquids, a sound alternative would be
liquid argon (LAr), since  $Z_{Ar} = 18$ and its density is $\sim$1.4~g/cc. A
LAr TPC  is a viable alternative to a LXe TPC: LAr TPC's with the fiducial
volume necessary for a Compton telescope have been built \cite{FArneodo:98}, LAr
is purified more easily than  LXe, LAr is relatively inexpensive and, last but
not least, LAr allows a better energy resolution than LXe \cite{EAprile:nim87}.

\section*{Conclusions}

In this work, stimulated by \cite{SEBoggs:nim03}, we have presented the
detection of polarization for Compton scattered \g-rays at 1.836~\MeV; data from
the LXeGRIT Compton telescope have been used. On the one hand, polarization has
been detected; on the other hand, it is clear to us that such an effect is small
when any high density, high $Z$ material (LXe in our case) is used as active
medium. Based on our experience with LXeGRIT, LAr is proposed as a promising
alternative.  

\section*{Acknowledgments}

This research was supported by NASA grant NAG5-5108 to the Columbia Astrophysics
Laboratory. 

\bibliography{reply-boggs}

\end{document}